\documentclass[12pt]{article}

\usepackage{amsmath}

\def\ltwid{\mathrel{\raise.3ex\hbox{$<$\kern-.75em\lower1ex\hbox{$\sim$}}}}
\def\gtwid{\mathrel{\raise.3ex\hbox{$>$\kern-.75em\lower1ex\hbox{$\sim$}}}}

\def\square{\kern1pt\vbox{\hrule height 1.2pt\hbox{\vrule width 1.2pt\hskip 3pt
   \vbox{\vskip 6pt}\hskip 3pt\vrule width 0.6pt}\hrule height 0.6pt}\kern1pt}
\def\overleftrightarrow#1{\vbox{\ialign{##\crcr
     $\leftrightarrow$\crcr\noalign{\kern-1pt\nointerlineskip}
     $\hfil\displaystyle{#1}\hfil$\crcr}}}
\def\dlim{\mathrel{\raise.8ex\hbox{${\scriptscriptstyle D = 4}$\kern-1.5em\lower1ex\hbox{$\longrightarrow$}}}}

\begin{document}

\begin{titlepage}

\begin{flushright}
CCTP-2012-16 \\ UFIFT-QG-12-07
\end{flushright}

\vskip 2cm

\begin{center}
{\bf Computing the Primordial Power Spectra Directly}
\end{center}

\vskip 1cm

\begin{center}
Maria G. Romania$^{*}$, N. C. Tsamis$^{\dagger}$
\end{center}

\begin{center}
Institute of Theoretical Physics \& Computational Physics \\
Department of Physics University of Crete \\
GR-710 03 Heraklion, HELLAS
\end{center}

\begin{center}
and
\end{center}

\begin{center}
R. P. Woodard$^{\ddagger}$
\end{center}

\begin{center}
Department of Physics, University of Florida \\
Gainesville, FL 32611, UNITED STATES
\end{center}

\vskip 1cm

\begin{center}
ABSTRACT
\end{center}
The tree order power spectra of primordial inflation depend upon the
norm-squared of mode functions which oscillate for early times and
then freeze in to constant values. We derive simple differential
equations for the power spectra, that avoid the need to numerically
simulate the physically irrelevant phases of the mode functions. We
also derive asymptotic expansions which should be valid until a few
e-foldings before first horizon crossing, thereby avoiding the need
to evolve mode functions from the ultraviolet over long periods of
inflation.

\begin{flushleft}
\end{flushleft}

\vskip 1cm

\begin{flushleft}
$^{*}$ e-mail: romania@physics.uoc.gr \\
$^{\dagger}$ e-mail: tsamis@physics.uoc.gr \\
$^{\ddagger}$ e-mail: woodard@phys.ufl.edu
\end{flushleft}

\end{titlepage}

\section{Introduction}\label{intro}

Cosmological perturbations from primordial inflation
\cite{Starobinsky:1979ty,Mukhanov:1981xt,Starobinsky:1982ee} have a
crucial importance for fundamental theory because they are first
quantum gravitational effect ever detected \cite{Mukhanov:2007zz,
Woodard:2009ns}, because they provide strong evidence for primordial
inflation \cite{Mukhanov:1990me,Liddle:1993fq,Lidsey:1995np}, and
because they set the initial conditions for structure formation in
cosmology \cite{Dodelson:2003ft,Mukhanov:2005sc,Weinberg:2008zzc}.
Two sorts of perturbations are produced during single-scalar
inflation: a scalar perturbation characterized by the field
$\zeta(t,\vec{x})$, and a tensor perturbation characterized by the
transverse-traceless field $h_{ij}(t,\vec{x})$. The scalar signal
has been imaged in the anisotropies of the cosmic microwave
background \cite{Komatsu:2010fb,Keisler:2011aw}, and by measuring
the matter power spectrum with large scale structure surveys
\cite{Reid:2009xm}. The tensor signal has not been imaged so far
\cite{Komatsu:2010fb,Keisler:2011aw} but strenuous efforts are
underway to detect it through the polarization of the cosmic
microwave background \cite{PLANCK,EBEX,SPIDER,PIPER}.

The two perturbation fields are defined on the homogeneous,
isotropic and spatially flat geometry characterized by scale factor
$a(t)$, with Hubble parameter $H(t)$ and slow roll parameters
$\epsilon(t)$ and $\eta(t)$,
\begin{equation}
ds^2 = -dt^2 + a^2(t) d\vec{x} \!\cdot\! d\vec{x} \; , \; H(t)
\equiv \frac{\dot{a}}{a} \; , \; \epsilon(t) \equiv
-\frac{\dot{H}}{H^2} \; , \; \eta(t) \equiv \epsilon -
\frac{\dot{\epsilon}}{2 H \epsilon} \; . \label{geometry}
\end{equation}
Only the quadratic parts of their Lagrangians are relevant to
current measurements (see \cite{Kahya:2010xh} for an explanation of
the full formalism),
\begin{eqnarray}
\mathcal{L}^{(2)}_{\zeta} & = & \frac{\epsilon a^3}{8 \pi G}
\Biggl\{ \dot{\zeta}^2 \!-\! \frac{\partial_k \zeta \partial_k
\zeta}{a^2} \Biggr\} \; , \label{Lzeta} \\
\mathcal{L}^{(2)}_{h} & = & \frac{a^3}{64 \pi G} \Biggl\{
\dot{h}_{ij} \dot{h}_{ij} \!-\! \frac{\partial_k h_{ij} \partial_k
h_{ij}}{a^2} \Biggr\} \; . \label{Lh}
\end{eqnarray}
The reported results
\cite{Komatsu:2010fb,Keisler:2011aw,Reid:2009xm} for the scalar and
tensor power spectra are consistent with evaluating the following
2-point correlators long after the first horizon crossing time $t_k$
such that $k = H(t_k) \, a(t_k)$,
\begin{eqnarray}
\Delta^2_{\mathcal{R}}(t,k) & \equiv & \frac{k^3}{2 \pi^2} \int \!\!
d^3x \, e^{-i \vec{k} \cdot \vec{x}} \Bigl\langle \Omega \Bigl\vert
\zeta(t,\vec{x}) \zeta(t,\vec{0}) \Bigr\vert \Omega \Bigr\rangle \;
, \label{DeltaR} \\
\Delta^2_{h}(t,k) & \equiv & \frac{k^3}{2 \pi^2} \int \!\! d^3x \,
e^{-i \vec{k} \cdot \vec{x}} \Bigl\langle \Omega \Bigl\vert
h_{ij}(t,\vec{x}) h_{ij}(t,\vec{0}) \Bigr\vert \Omega \Bigr\rangle
\; . \label{Deltah}
\end{eqnarray}

At tree order the correlators (\ref{DeltaR}-\ref{Deltah}) can be
expressed in terms of the scalar mode function $v(t,k)$ and its
tensor cousin $u(t,k)$,
\begin{eqnarray}
\Delta^2_{\mathcal{R}}(t,k) & = & \frac{k^3}{2 \pi^2} \times 4\pi G
\times \vert v(t,k) \vert^2 + O(G^2) \; , \label{treezeta} \\
\Delta^2_{h}(t,k) & = & \frac{k^3}{2 \pi^2} \times 32 \pi G \times 2
\times \vert u(t,k) \vert^2 + O(G^2) \; . \label{treeh}
\end{eqnarray}
The relevant mode equations and normalization conditions are,
\begin{eqnarray}
\ddot{v} + \Bigl(3 H + \frac{\dot{\epsilon}}{\epsilon}\Bigr) \dot{v}
+ \frac{k^2}{a^2} v = 0 \quad & , & \quad W_v \equiv v \dot{v}^* -
\dot{v} v^* = \frac{i}{\epsilon a^3} \; , \label{veqns} \\
\ddot{u} + 3 H \dot{u} + \frac{k^2}{a^2} u = 0 \quad & , & \quad W_u
\equiv u \dot{u}^* - \dot{u} u^* = \frac{i}{a^3} \; . \label{ueqns}
\end{eqnarray}
The initial conditions for the mode functions (which correspond to
the choice of Bunch-Davies vacuum) derive from the WKB solutions in
the far ultraviolet (that is, for $k \gg H(t) a(t)$),
\begin{eqnarray}
v(t,k) & \longrightarrow & \frac{\exp[-ik \int^t dt'/a(t')]}{\sqrt{2
k a^2(t) \epsilon(t)}} \Biggl\{ 1 + O\Bigl(\frac{H a}{k}\Bigr) \Biggr\}
\; , \label{initialzeta} \\
u(t,k) & \longrightarrow & \frac{\exp[-ik \int^t dt'/a(t')]}{\sqrt{2
k a^2(t)}} \Biggl\{1 + O\Bigl(\frac{Ha}{k}\Bigr) \Biggr\} \; .
\label{initialh}
\end{eqnarray}
As is evident from these asymptotic forms, both mode functions
oscillate and fall off in the far ultraviolet. They become
approximately constant near the time of first horizon crossing
$t_k$. (One can infer the existence of constant solutions quite
generally when the $\frac{k^2}{a^2(t)}$ terms in equations
(\ref{veqns}-\ref{ueqns}) become irrelevant.) It is these constant
amplitudes which determine the crucial theoretical predictions for
the power spectra through relations (\ref{treezeta}-\ref{treeh}).

One of the frustrating things about this formalism is the need to
employ approximation techniques, even for evaluating tree order
formulae such as (\ref{treezeta}-\ref{treeh}), because equations
(\ref{veqns}-\ref{ueqns}) for the mode functions cannot be solved
analytically for general scale factor $a(t)$
\cite{Tsamis:2002qk,Tsamis:2003zs}. Examples of such approximation
techniques include \cite{Stewart:1993bc,Wang:1997cw,Martin:1999wa}:
\begin{itemize}
\item{Assuming the slow roll parameter $\epsilon(t)$ is constant;}
\item{Matching the leading ultraviolet and infrared forms at $t = t_k$; and}
\item{Employing the full WKB solutions.}
\end{itemize}
When greater accuracy is needed, one must resort to numerical
evolution of (\ref{veqns}-\ref{ueqns}) from the known initial
conditions (\ref{initialzeta}-\ref{initialh}) until well past the
time of first horizon crossing \cite{Wang:1997cw}. Excellent
numerical solution techniques exist
\cite{Easther:2010qz,Mortonson:2010er,Easther:2011yq}, but a large
fraction of their power is wasted on reproducing the oscillations of
the mode functions, which contribute nothing to the power spectra.
This can be especially time-consuming for models in which there is
an extended phase of inflation, or when scanning properties of
classes of models.

A more effective technique was developed recently for the tensor
power spectrum \cite{Romania:2011ez} in order to work out the
gravitational wave signal from a novel model of inflation in which
the Hubble parameter oscillates for a brief time
\cite{Tsamis:2009ja,Romania:2010zq}. The key to the new technique is
to convert relations (\ref{ueqns}) into an equation for the
norm-squared of the mode function, $M(t,k) \equiv \vert u(t,k)
\vert^2$. This is the quantity that actually enters the tensor power
spectrum (\ref{treeh}), and it has a much more sedate evolution than
the mode function. Evolving $M(t,k)$ avoids the wasted effort of
numerically simulating the oscillations of $u(t,k)$. Further, an
excellent asymptotic expansion can be derived for $M(t,k)$ which
must be valid until only a few e-foldings before first horizon
crossing, no matter how long inflation persists
\cite{Romania:2011ez}. It is therefore only necessary to numerically
evolve $M(t,k)$ for a few e-foldings, from just before $t_k$ until
it becomes constant.

The purpose of this paper is to extend this more effective technique
to the scalar mode functions using a trick \cite{Tsamis:2003zs} for
converting solutions to $u(t,k)$ into solutions for $v(t,k)$. To
further simplify the formalism, we express our results in terms of
the tree order power spectra (\ref{treezeta}-\ref{treeh}) directly,
without ever mentioning the mode functions. Section \ref{eqn}
derives the key equations for $\Delta^2_{\mathcal{R}}(t,k)$ and
$\Delta^2_{h}(t,k)$, and section \ref{expand} gives their asymptotic
expansions. Our discussion comprises section \ref{discuss}.

\section{Differential Equations for $\Delta^2_{\mathcal{R}}$ and
$\Delta^2_{h}$}\label{eqn}

We begin by reviewing how one derives an equation for $M(t,k) \equiv
\vert u(t,k) \vert^2$. The first step is to write out the first and
second time derivatives,
\begin{eqnarray}
\dot{M}(t,k) & = & \dot{u} u^* + u \dot{u}^* \; , \label{Mdot} \\
\ddot{M}(t,k) & = & \ddot{u} u^* + 2 \dot{u} \dot{u}^* + u
\ddot{u}^* \; . \label{Mddot}
\end{eqnarray}
One next eliminates $\ddot{u}$ and $\ddot{u}^*$ in (\ref{Mddot}) by
using (\ref{ueqns}), and then recognizing factors of $M$ and
$\dot{M}$,
\begin{equation}
\ddot{M}(t,k) = -3 H \dot{M} - \frac{2 k^2}{a^2} M + 2 \dot{u}
\dot{u} \; . \label{Mstep2}
\end{equation}
The final factor involving $\dot{u} \dot{u}^*$ can be expressed in
terms of $M$ by subtracting the square of the Wronskian $W_u$ in
expression (\ref{ueqns}) from the square of $\dot{M}$ in expression
(\ref{Mdot}),
\begin{equation}
\dot{M}^2 - W^2_u = 2 M \times 2 \dot{u} \dot{u}^* = \dot{M}^2 +
\frac1{a^6} \qquad \Longrightarrow \qquad 2 \dot{u} \dot{u}^* =
\frac{\dot{M}^2}{2 M} + \frac1{2 M a^6} \; . \label{dotuu*}
\end{equation}
Substituting (\ref{dotuu*}) in relation (\ref{Mstep2}) gives the
desired equation for $M(t,k)$,
\begin{equation}
\ddot{M} + 3 H \dot{M} + \frac{2 k^2}{a^2} M = \frac{ \dot{M}^2}{2
M} + \frac1{2 M a^6} \; . \label{Meqn}
\end{equation}

Deriving an equation for $N(t,k) \equiv \vert v(t,k)\vert^2$ entails
the same first step,
\begin{eqnarray}
\dot{N}(t,k) & = & \dot{v} v^* + v \dot{v}^* \; , \label{Ndot} \\
\ddot{N}(t,k) & = & \ddot{v} v^* + 2 \dot{v} \dot{v}^* + v
\ddot{v}^* \; . \label{Nddot}
\end{eqnarray}
We next employ expression (\ref{veqns}) to eliminate $\ddot{v}$ and
$\ddot{v}^*$ in favor of $N$ and $\dot{N}$,
\begin{equation}
\ddot{N} = -\Bigl(3 H + \frac{\dot{\epsilon}}{\epsilon}\Bigr)
\dot{N} - \frac{2 k^2}{a^2} N + 2 \dot{v} \dot{v}^* \; .
\label{Nstep2}
\end{equation}
Subtracting the squares of the $\dot{N}$ from (\ref{Ndot}) and the
Wronskian $W_v$ in equation (\ref{veqns}) gives $\dot{v} \dot{v}^*$,
\begin{equation}
\dot{N}^2 - W^2_v = 2 N \times 2 \dot{v} \dot{v}^* = \dot{N}^2 +
\frac1{\epsilon^2 a^6} \qquad \Longrightarrow \qquad 2 \dot{v}
\dot{v}^* = \frac{\dot{N}^2}{2 N} + \frac1{2 N \epsilon^2 a^6} \; .
\label{dotvv*}
\end{equation}
And the final equation for $N(t,k)$ comes from substituting
(\ref{dotvv*}) in (\ref{Nstep2}),
\begin{equation}
\ddot{N} + \Bigl(3 H + \frac{\dot{\epsilon}}{\epsilon}\Bigr) \dot{N}
+ \frac{2 k^2}{a^2} N = \frac{\dot{N}^2}{2 N} + \frac1{2 N
\epsilon^2 a^6} \; . \label{Neqn}
\end{equation}

Because equations (\ref{Meqn}) and (\ref{Neqn}) are nonlinear, they
incorporate the Wronskian normalization as well as the evolution
equations for the mode functions. Converting them to equations for
$\Delta^2_{\mathcal{R}}(t,k)$ and $\Delta^2_{h}(t,k)$ requires only
changing the coefficients of the final terms,
\begin{eqnarray}
\ddot{\Delta}^2_{\mathcal{R}} + \Bigl(3 H +
\frac{\dot{\epsilon}}{\epsilon} \Bigr) \dot{\Delta}^2_{\mathcal{R}}
+ \frac{2 k^2}{a^2} \Delta^2_{\mathcal{R}} =
\frac{(\dot{\Delta}^2_{\mathcal{R}})^2}{\Delta^2_{\mathcal{R}}} +
\frac{2 G^2 k^6}{\pi^2 \epsilon^2 a^6}
\frac1{\Delta^2_{\mathcal{R}}} \; , \label{scalareqn} \\
\ddot{\Delta}^2_{h} + 3 H \dot{\Delta}^2_{h} + \frac{2 k^2}{a^2}
\Delta^2_{h} = \frac{(\dot{\Delta}^2_{h})^2}{\Delta^2_{h}} +
\frac{2^9 G^2 k^6}{\pi^2 a^6} \frac1{\Delta^2_{h}} \; .
\label{tensoreqn}
\end{eqnarray}

\section{Asymptotic Expansions for $\Delta^2_{\mathcal{R}}$ and
$\Delta^2_{h}$} \label{expand}

Equations (\ref{scalareqn})-\ref{tensoreqn}) require initial values
for the power spectra and their first time derivatives. Of course
these derive from the ultraviolet limits
(\ref{initialzeta}-\ref{initialh}), which imply the zeroth order
terms for an expansion in powers of $(H a/k)^2$,
\begin{eqnarray}
\Delta^2_{\mathcal{R}}(t,k) & \longrightarrow & \frac{G k^2}{\pi
\epsilon a^2} \Biggl\{ 1 + O\Bigl( \frac{H^2 a^2}{k^2}\Bigr)
\Biggr\} \; , \label{zeta0term} \\
\Delta^2_{h}(t,k) & \longrightarrow & \frac{16 G k^2}{\pi a^2}
\Biggl\{ 1 + O\Bigl( \frac{H^2 a^2}{k^2} \Bigr) \Biggr\} \; .
\label{h0term}
\end{eqnarray}
We shall develop the expansion for $\Delta^2_{h}(t,k)$ to second
order, and then employ a trick for converting tensor mode functions
to scalar ones \cite{Tsamis:2003zs} to obtain the corresponding
expansion of $\Delta^2_{\mathcal{R}}(t,k)$. These expansions are so
accurate that there seems little point to numerically evolving for
anything except the last few e-foldings before first horizon
crossing.

From the asymptotic limit (\ref{h0term}) one can see that the
``big'' terms in the evolution equation (\ref{tensoreqn}) are those
with no derivatives. It is best to segregate these terms,
\begin{equation}
\Delta^2_{h} - \Bigl( \frac{16 \pi G k^2}{\pi a^2} \Bigr)^2
\frac1{\Delta^2_{h}} = \frac{a^2}{2 k^2} \Biggl[-\ddot{\Delta}^2_{h}
- 3 H \dot{\Delta}^2_{h} + \frac{ (\dot{\Delta}^2_{h})^2}{2
\Delta^2_{h}} \Biggr] \; . \label{perteqn}
\end{equation}
The desired expansion is now easy to read off \cite{Romania:2011ez},
\begin{eqnarray}
\lefteqn{ \Delta^2_{h}(t,k) = \frac{16 G k^2}{\pi a^2} \Biggl\{ 1 +
\Bigl[1 - \frac12 \epsilon\Bigr] \Bigl( \frac{H a}{k}\Bigr)^2 }
\nonumber \\
& & \hspace{.5cm} + \Biggl[ \frac94 \epsilon \!-\! \frac{21}8
\epsilon^2 \!+\! \frac34 \epsilon^3 \!+\! \Bigl( \frac74 \!-\!
\frac34 \epsilon\Bigr) \frac{\dot{\epsilon}}{H} \!+\! \frac18
\frac{\ddot{\epsilon}}{H^2} \Biggr] \Bigl( \frac{H a}{k}\Bigr)^4 +
O\Bigl( \frac{H^6 a^6}{k^6} \Bigr) \Biggr\} \; . \qquad
\label{tensorexp}
\end{eqnarray}
Note that this expansion is in powers of $(\frac{H a}{k})^2$, rather
than the expansion in powers of $\frac{H a}{k}$ one gets for the
mode function. The better convergence for small $\frac{H a}{k}$ is
one more indication of the superiority of our method.

The trick for converting $\Delta^2_{h}(t,k)$ into
$\Delta^2_{\mathcal{R}}(t,k)$ involves simultaneously changing the
scale factor and the meaning of time \cite{Tsamis:2003zs},
\begin{eqnarray}
a(t) & \longrightarrow & \sqrt{\epsilon(t)} \times a(t) \; ,
\label{newa} \\
\frac{\partial}{\partial t} & \longrightarrow &
\frac1{\sqrt{\epsilon(t)}} \times \frac{\partial}{\partial t} \; .
\label{newdt}
\end{eqnarray}
One can see that these two replacements convert the $u(t,k)$ mode
relations (\ref{ueqns}) into the relations (\ref{veqns}) for
$v(t,k)$. Replacements (\ref{newa}-\ref{newdt}) also convert the
$\Delta^2_{h}(t,k)$ equation (\ref{tensoreqn}) into expression
(\ref{scalareqn}) for $\Delta^2_{\mathcal{R}}(t,k)$. We could have
used them to derive that equation had the straightforward derivation
not been so trivial.

To convert the asymptotic expansion (\ref{tensorexp}) into an
expansion for $\Delta^2_{\mathcal{R}}(t,k)$ we first note that the
replacements (\ref{newa}-\ref{newdt}) imply the following
geometrical replacements,
\begin{eqnarray}
H & \longrightarrow & \frac{H}{\sqrt{\epsilon}} \times \Bigl[ 1
\!+\! \epsilon \!-\! \eta \Bigr] \equiv \frac{H}{\sqrt{\epsilon}}
\times D \; , \label{newH} \\
\epsilon & \longrightarrow & \frac{ (1 \!+\! \epsilon \!-\! \eta) (2
\epsilon \!-\! \eta) \!-\! (\frac{\dot{\epsilon} - \dot{\eta}}{H})}{
(1 \!+\! \epsilon \!-\! \eta)^2 } \equiv \frac{N}{D^2} \; , \label{neweps} \\
\frac{\dot{\epsilon}}{H} & \longrightarrow & \frac{D \dot{N} \!-\! 2
N \dot{D}}{H D^4} \; , \label{newdoteps} \\
\frac{\ddot{\epsilon}}{H^2} & \longrightarrow & \frac{D^2 \ddot{N}
\!-\! 4 D \dot{N} \dot{D} \!-\! 2 D N \ddot{D} \!+\! 6 N
\dot{D}^2}{H^2 D^6} - \frac{ (\epsilon \!-\! \eta) (D \dot{N} \!-\!
2 N \dot{D})}{H D^5} \; . \label{newddoteps}
\end{eqnarray}
After many tedious manipulations the result is,
\begin{eqnarray}
\lefteqn{ \Delta^2_{\mathcal{R}}(t,k) = \frac{G k^2}{\pi \epsilon
a^2} \Biggl\{ 1 + \Biggl[ 1 \!-\! \frac12 (1 \!+\! \epsilon \!-\!
\eta) (2 \epsilon \!-\! \eta) - \Bigl( \frac{\dot{\epsilon} \!-\!
\dot{\eta}}{2 H} \Bigr) \Biggr] \Bigl( \frac{H a}{k} \Bigr)^2 }
\nonumber \\
& & \hspace{-.7cm} + \Biggl[ \frac38 (1 \!+\! \epsilon \!-\! \eta)
(3 \!-\! \epsilon \!-\! \eta) (2 \!-\! \eta) (2 \epsilon \!-\! \eta)
\!+\! \Bigl[ \frac54 \!+\! \frac92 \epsilon \!-\! \frac{17}4 \eta
\!-\! \frac78 \epsilon^2 \!-\! \frac32 \epsilon \eta \!+\!
\frac{13}8 \eta^2 \Bigr] \frac{\dot{\epsilon}}{H} \nonumber \\
& & \hspace{-.7cm} + \Bigl[ \frac12 \!-\! \frac{21}8 \epsilon \!+\!
\frac{13}8 \eta \!+\! \epsilon^2 \!+\! \frac12 \epsilon \eta \!-\!
\frac34 \eta^2 \Bigr] \frac{\dot{\eta}}{H} \!+\! \frac74 \Bigl(
\frac{\dot{\epsilon}}{H}\Bigr)^2 \!-\! \frac{25}8
\Bigl(\frac{\dot{\epsilon}}{H} \Bigr) \Bigl(
\frac{\dot{\eta}}{H}\Bigr) \!+\! \frac{11}8 \Bigl(
\frac{\dot{\eta}}{H} \Bigr)^2 \nonumber \\
& & \hspace{-.7cm} - \Bigl[ \frac32 \!+\! \frac38 \epsilon \!-\!
\frac34 \eta \Bigr] \frac{\ddot{\epsilon}}{H^2} \!+\! \Bigl[
\frac{13}8 \!+\! \frac{\epsilon}2 \!-\! \frac78 \eta\Bigr]
\frac{\ddot{\eta}}{H^2} \!-\! \Bigl( \frac{\dddot{\epsilon} \!-\!
\dddot{\eta}}{8 H^3} \Bigr) \! \Biggr] \! \Bigl( \frac{Ha}{k}
\Bigr)^4 \!\!\!+\! O\Bigl( \frac{H^6 a^6}{k^6} \Bigr) \! \Biggr\} .
\label{scalarexp} \quad
\end{eqnarray}
Unless $\epsilon(t)$ or $\eta(t)$ is changing rapidly on the Hubble
scale, the fractional error in (\ref{scalarexp}) should be of order
$(\frac{Ha}{k})^6$. Had we used this formula to estimate
$\Delta^2_{\mathcal{R}}(t,k)$ for $t$ as close as one e-folding
before first horizon crossing, the fractional error would only be
about $2.5 \times 10^{-3}$; for two e-foldings that would fall to
about $6.1 \times 10^{-6}$; and the fractional error would only be
about $1.5 \times 10^{-8}$ at three e-foldings before horizon
crossing. This is comparable to what loop corrections might give
\cite{Miao:2012xc}.

\section{Discussion}\label{discuss}

We have derived nonlinear, second order differential equations
(\ref{scalareqn}) and (\ref{tensoreqn}) for the correlators
$\Delta^2_{\mathcal{R}}(t,k)$ and $\Delta^2_{h}(t,k)$ whose late
time limits give the scalar and tensor power spectra at tree order.
Numerically evolving these correlators is more economical than
evolving the mode functions (\ref{veqns}-\ref{ueqns}) because one no
longer has to keep track of the phases which drop out of the power
spectra. This economy shows up in the asymptotic expansions
(\ref{scalarexp}) and (\ref{tensorexp}) pertinent to the ultraviolet
regime of $k \gg H(t) \, a(t)$. Our expansions for the correlators
are in powers of the small parameter $(\frac{Ha}{k})^2$, whereas the
analogous expansions for the mode functions are only in powers of
$\frac{Ha}{k}$. In fact our asymptotic expansions
(\ref{scalarexp}-\ref{tensorexp}) should be so accurate that they
could be used to provide the initial value data for
$\Delta^2_{\mathcal{R}}(t,k)$ and $\Delta^2_{h}(t,k)$ only a few
e-foldings before the time $t_k$ of first horizon crossing.

We have employed ``Hubble parametrization'', in which the scale
factor $a(t)$ is assumed known and results are expressed in terms of
background geometrical quantities $H(t)$, $\epsilon(t)$ and
$\eta(t)$ defined in expression (\ref{geometry}). This is especially
useful for the tensor power spectrum because gravitational waves
depend only on the background geometry, no matter what caused it.
Many people prefer ``potential parametrization'', in which the
scalar potential $V(\phi)$ is assumed known (with the field starting
at rest from some fiducial value) and results are expressed as
derivatives of the potential. Approximate conversion formulae are,
\begin{equation}
H^2 \simeq \frac83 \pi G V \quad , \quad \epsilon \simeq \frac1{16
\pi G} \Bigl( \frac{V'}{V} \Bigr)^2 \quad , \quad \eta \simeq
\frac1{16 \pi G} \Biggl[ 2\frac{V''}{V} \!-\! \Bigl( \frac{V'}{V}
\Bigr)^2 \Biggr] \; .
\end{equation}

Finally, we note that the same technology can be applied to derive
an equation for the power spectrum of any field whose plane wave
mode functions obey the Mukhanov equation \cite{Tsamis:2003zs}. Of
course the associated asymptotic expansion is also easy to develop.
Both the equation and its solution in the regime of $ k \gg H(t) \,
a(t)$ can be obtained from the analogous tensor relations,
(\ref{tensoreqn}) and (\ref{tensorexp}), by changing the scale
factor and the time as we did in equations (\ref{newa}-\ref{newdt}).

\vskip .5cm

\centerline{\bf Acknowledgements}

We are grateful to W. Kinney and M. Sasaki for conversations on this
subject. This work was partially supported by European Union program
Thalis ESF/NSRF 2007-2013, by European Union Grant
FP-7-REGPOT-2008-1-CreteHEPCosmo-228644, by NSF grant PHY-1205591,
and by the Institute for Fundamental Theory at the University of
Florida.

\end{document}